\documentclass{aa}
\usepackage{epsf}

\begin{document}

% *************************************************************************
%                             ABBREVIATIONS
% *************************************************************************
\def\pFn{p_{\raise-0.3ex\hbox{{\scriptsize F$\!$\raise-0.03ex\hbox{\rm n}}}}
}  % p_Fn
\def\pFp{p_{\raise-0.3ex\hbox{{\scriptsize F$\!$\raise-0.03ex\hbox{\rm p}}}}
}  % p_Fp
\def\pFe{p_{\raise-0.3ex\hbox{{\scriptsize F$\!$\raise-0.03ex\hbox{\rm e}}}}
}  % p_Fe
\def\pFmu{p_{\raise-0.3ex\hbox{{\scriptsize F$\!$\raise-0.03ex\hbox{\rm
$\mu$}}}} }  % p_Fe
\def\m@th{\mathsurround=0pt }
\def\eqalign#1{\null\,\vcenter{\openup1\jot \m@th
   \ialign{\strut$\displaystyle{##}$&$\displaystyle{{}##}$\hfil
   \crcr#1\crcr}}\,}
\newcommand{\vp}{\mbox{\boldmath $p$}}         % vector p (momentum)
\newcommand{\vS}{\mbox{\boldmath $S$}}
\newcommand{\vP}{\mbox{\boldmath $P$}}

\newcommand{\vk}{\mbox{\boldmath $k$}}         % vector k (momentum)
\newcommand{\xixi}{\mbox{\boldmath $\xi$}}         % vector kappa (momentum)
\newcommand{\vq}{\mbox{\boldmath $q$}}         % vector q (momentum)
\newcommand{\vr}{\mbox{\boldmath $r$}}         % vector r (momentum)

\newcommand{\om}{\mbox{$\omega$}}              % \omega
\newcommand{\Om}{\mbox{$\Omega$}}              % \Omega
\newcommand{\Th}{\mbox{$\Theta$}}              % \Theta
\newcommand{\ph}{\mbox{$\varphi$}}             % \varphi
\newcommand{\del}{\mbox{$\delta$}}             % \delta
\newcommand{\Del}{\mbox{$\Delta$}}             % \Delta
\newcommand{\lam}{\mbox{$\lambda$}}            % \lambda
\newcommand{\Lam}{\mbox{$\Lambda$}}            % \Lambda
\newcommand{\ep}{\mbox{$\varepsilon$}}         % \varepsilon
\newcommand{\ka}{\mbox{$\kappa$}}              % \kappa
\newcommand{\dd}{\mbox{d}}                     % d - differential
\newcommand{\vect}[1]{\bf #1}                % vector (Äŧ¬å¨ ¡§´Áµ¨)
\newcommand{\vtr}[1]{\mbox{\boldmath $#1$}}  % vector (Äŧ¬å¨ ¬ê¦'µ¬¬å¨)
\newcommand{\vF}{\mbox{$v_{\mbox{\raisebox{-0.3ex}{\scriptsize F}}}$}}  %v_F
\newcommand{\pF}{\mbox{$p_{\mbox{\raisebox{-0.3ex}{\scriptsize F}}}$}}  %p_F
\newcommand{\kF}{\mbox{$k_{\rm F}$}}           % k_F
\newcommand{\kTF}{\mbox{$k_{\rm TF}$}}         % k_TF
\newcommand{\kB}{\mbox{$k_{\rm B}$}}           % k_B
\newcommand{\tn}{\mbox{$T_{{\rm c}n}$}}        % Tcn (crit.tem-re of neutrons)
\newcommand{\tp}{\mbox{$T_{{\rm c}p}$}}        % Tcp (crit.tem-re of protons)
\newcommand{\te}{\mbox{$T_{eff}$}}             % T effective
\newcommand{\ex}{\mbox{\rm e}}                 % exponent (roman type)
\newcommand{\rate}{\mbox{${\rm erg~cm^{-3}~s^{-1}}$}}
\newcommand{\mur}{\raisebox{0.2ex}{\mbox{\scriptsize (í)}}} %--------------%
\newcommand{\Mn}{\raisebox{0.2ex}{\mbox{\scriptsize (í{\it n\/})}}}        %
\newcommand{\Mp}{\raisebox{0.2ex}{\mbox{\scriptsize (í{\it p\/})}}}        %
\newcommand{\MN}{\raisebox{0.2ex}{\mbox{\scriptsize (í{\it N\/})}}}        %

% *************************************************************************
%                                 TITLE
% *************************************************************************
\title{ \bf Enhanced cooling of neutron stars via
           Cooper-pairing neutrino emission
}

\author{ M.~E.\ Gusakov \inst{1},  
         A.~D.\ Kaminker \inst{1}
\and	 
         D.G.~Yakovlev \inst{1}
\and  
         O.~Y.\ Gnedin \inst{2}	 
	   }
\institute{
         Ioffe Physical Technical Institute,
         Politekhnicheskaya 26, 194021 St.~Petersburg, Russia,
        {\it gusakov@astro.ioffe.ru; kam@astro.ioffe.ru; yak@astro.ioffe.ru } 
        \and
        Space Telescope Science Institute,
        3700 San Martin Drive, Baltimore, MD 21218, USA,
	{\it ognedin@stsci.edu}
}
\offprints{M.E.\ Gusakov}
\date{Received xx xxxxx 2004 / Accepted xx xxxxx 2004}
% **********************************************************************
\abstract
{
We simulate cooling of superfluid neutron stars with
nucleon cores where
direct Urca process is forbidden. We adopt density
dependent critical temperatures $T_{\rm cp}(\rho)$ and
$T_{\rm cn}(\rho)$ of singlet-state proton and triplet-state
neutron pairing in a stellar core and consider a strong
proton pairing (with maximum $T_{\rm cp}^{\rm max} \ga
5 \times 10^9$ K) and a moderate neutron pairing
($T_{\rm cn}^{\rm max} \sim 6 \times 10^8$ K).
When the internal stellar temperature $T$ falls below
$T_{\rm cn}^{\rm max}$, the neutrino luminosity $L_{\rm CP}$
due to Cooper pairing of neutrons  behaves
$\propto T^8$, just as that produced by modified
Urca process (in a non-superfluid star)
but is higher by about two orders of magnitude.
In this case the Cooper-pairing neutrino emission acts
like an enhanced cooling agent. By tuning the density
dependence $T_{\rm cn}(\rho)$ we can explain observations
of cooling isolated neutron stars in the scenario in which direct Urca
process or similar process in kaon/pion condensed or
quark matter are absent.

\keywords{Stars: neutron -- dense matter}
}
\titlerunning{Enhanced cooling of neutron stars}
\authorrunning{M.~E.~Gusakov, A.~D.~Kaminker, D.~G.~Yakovlev, O.~Y.~Gnedin}
\maketitle

%%%%%%%%%%%%%%%%%%%%%%%%%%%%%%%%%%%%%%%%%%%%%%%%%%%%%%%%%%%%%%%%%%%%%%%%%%%%%
%**************** Section 1 ******************************
\section{Introduction}
\label{introduction}
%%%%%%%%%%%%%%%%%%%%%%%%%%%%%%%%%%%%%%%%%%%%%%%%%%%%%%%%%%%%%

Thanks to {\it Chandra} and {\it XMM-Newton} missions,
there is a great progress in observations of thermal
radiation emergent from the surfaces of isolated
(cooling) middle-aged neutron stars (e.g., Pavlov \& Zavlin
\cite{pz03}). A comparison of these data with theoretical
models of cooling neutron stars gives a method
to constrain  (still poorly known)
fundamental properties of supranuclear
matter in neutron-star cores, such as the composition
and equation of state of the matter and its
superfluid properties.

So far, the observations can be explained by a number
of vastly different
theoretical models (e.g., 
Page \cite{page98a,page98b},
Tsuruta et al.\ \cite{tsurutaetal02},
Khodel et al.\ \cite{khodeletal04},
Blaschke et al.\ \cite{blaschke},
Yakovlev \& Pethick \cite{yp04}, and references therein).
Particularly, one can employ the 
simplest models of neutron stars with the cores
composed of nucleons (or nucleons/hyperons), or
containing pion condensates, kaon condensates or quarks.
The simplest model of a non-superfluid nucleon core
which cools via modified Urca process of neutrino emission
(without any powerful direct Urca process)
cannot explain the observations: some neutron
stars (e.g., PSR B1055--52) turn out to be much warmer, while others
(e.g., the Vela pulsar) are much colder than those given by
this model. Warmer stars can be explained
(Kaminker et al.\ \cite{khy01}) assuming a
strong proton superfluidity in the core: such a
superfluidity suppresses modified Urca process and
slows down the cooling. However colder stars
require some cooling mechanism which is faster than
the modified Urca process.

Explanations of observations of colder stars
presented in the literature invoke usually either
a powerful direct Urca process in nucleon (or nucleon/hyperon)
matter or similar processes in kaon-condensed,
pion-condensed, quark matter in the inner cores of massive
neutron stars.

In this paper we present a new scenario of neutron star
cooling. We adopt the simplest model equation of state
of supranuclear matter in neutron star cores
(Douchin \& Haensel \cite{dh01}) involving
only nucleons, electrons and muons. This equation of
state forbids direct Urca process in all stable neutron
stars. We will show that the enhanced cooling
required to explain colder isolated neutron stars
can be produced by neutrino emission due to a moderately
strong triplet-state pairing of neutrons.
This new interpretation is possible only for
a specific density dependence of the critical
temperature of neutron pairing.

In the next section we outline the observational basis;
the cooling scenario is given afterwards. 

%%%%%%%%%%%%%%%%%%%%%%%%%%%%%%%%%%%%%%%%%%%%%%%%%%%%%%%%%%%%%%%%%%%%%%%%%%%%
\section{Observations}
\label{observations}
%%%%%%%%%%%%%%%%%%%%%%%%%%%%%%%%%%%%%%%%%%%%%%%%%%%%%%%%%%%%%%%%%%%%%%%%%%%%

Table 1 summarize observations of isolated (cooling)
middle-aged ($10^3 \la t \la 10^6$ yr) neutron stars,
whose thermal surface radiation has been detected (or constrained).
We present the estimated stellar ages $t$ and
effective surface temperatures $T_{\rm s}^\infty$
(as detected by a distant observer).

Two young objects, RX J0822--4300 and 1E 1207.4--5209, 
are radio-quiet neutron stars in supernova remnants; 
RX J1856.4--3754 and RX J0720.4--3125 are also radio-quiet 
neutron stars. Other objects --- the Crab and the 
Vela pulsars, PSR B1706--44, 
PSR J0538+2817, Geminga, and PSR B1055--52 ---
are observed as radio pulsars. 

\renewcommand{\arraystretch}{1.2}
\begin{table*}[!t]   % "*" ignores the twocolumn-format if adopted
\caption[]{Observational limits on surface temperatures of isolated
neutron stars}
\label{tab:observ}
\begin{center}
\begin{tabular}{|| l | c | c | c | l ||}
\hline
\hline
Source & $t$ [kyr] & $T_{\rm s}^\infty$ [MK] &  Confid.\  & References   \\
\hline
\hline
PSR J0205+6449     & 0.82    & $<$1.1$^{~b)}$  & --     & 
Slane et al.\ (\cite{slane02})    \\
Crab               &    1    & $<$2.0$^{~b)}$   &  99.7\%    & 
Weisskopf et al.\ (\cite{weisskopf04})  \\
RX J0822--4300     & 2--5    & 1.6--1.9$^{~a)}$ & 90\% & 
Zavlin et al.\ (\cite{ztp99})   \\
%                                              Zavlin, Truemper, Pavlov 1999
1E 1207.4--5209        & 3--20 & 1.4--1.9$^{~a)}$ & 90\% & 
Zavlin et al.\ (\cite{zps03}) \\
%                                       Zavlin, Pavlov, Truemper 1998
Vela               & 11--25  & 0.65--0.71$^{~a)}$ & 68\% & 
Pavlov et al.\ (\cite{pavlovetal01})\\
%                                Pavlov, Zavlin,  Sanwal 2003 Bad Honnef
PSR B1706--44       & $\sim$17 & 0.82$^{+0.01}_{-0.34}$$^{~a)}$ & 68\% &
McGowan et al. (\cite{mcgowanetal04}) \\
%                                Pavlov, Zavlin,  Sanwal 2003 Bad Honnef
PSR J0538+2817       & $30 \pm 4$ & $\sim 0.87$$^{~a)}$ & -- &
Zavlin \& Pavlov (\cite{zp03}) \\
%PSR B0656+14       &  $\sim$110 & 0.53$^{+0.04}_{-0.03}$$^{~a)}$  & --  &
%Anderson et al.\ (1993) \\
Geminga            & $\sim$340 & $\sim 0.5$$^{~b)}$ & 90\% &
Zavlin \& Pavlov (\cite{zp03}) \\
RX~J1856.4--3754     & $\sim$500 & $<$0.65  & -- & see text \\
PSR~B1055--52       & $\sim$540 & $\sim$0.75$^{~b)}$ & -- &
Pavlov \& Zavlin (\cite{pz03})  \\
RX J0720.4--3125   & $\sim 1300$ & $\sim 0.51$$^{~a)}$  & -- &
Motch et al.\ (\cite{motchetal03}) \\
\hline
\end{tabular}
\end{center}
\small{
$^{a)}$ Inferred using a hydrogen atmosphere model\\
$^{b)}$ Inferred using the black-body spectrum\\
}
\end{table*}
\renewcommand{\arraystretch}{1.0}

RX J0205+6449 and the Crab pulsar are associated
with historical supernovae and their ages are certain.
For RX J0822--4300, we take 
the age of the host supernova remnant, Puppis A.
As can be deduced, e.g., from a discussion in Arendt et al.\ 
(\cite{adp91}),
its age ranges from 2 to 5 kyr; the central value
is $t = 3.7$ kyr (Winkler et al.\ \cite{winkler88}).
For 1E 1207.4--5209, we also adopt the age of 
the host supernova remnant (G296.5+10). 
According to Roger et al.\ (\cite{roger88}), it is $t \sim 3-20$ kyr.
For the Vela pulsar, we take the age interval from the 
standard characteristic spindown age of the pulsar
to the characteristic age 
corrected due to the pulsar glitching behaviour
(Lyne et al.\ \cite{lyne96}).
The age of PSR J0538+2817,
$t=(30 \pm 4)$ kyr, was estimated by Kramer et al.\ (\cite{kramer03})
from the measurements of the pulsar proper motion 
relative to the center of the host supernova remnant, S147.
The age of RX J1856.4--3754 has been revised recently by Walter \&
Lattimer (\cite{wl02}) from the kinematical reasons.
Following these authors we take the central
value $t= 500$ kyr and choose such an error-bar of $t$
to clearly distinguish the revised value from the value
$t=900$ kyr reported previously by Walter (\cite{walter01})
on the basis of less accurate parallax measurement.
The characteristic age of RX J0720.4--3125 has been
estimated by Zane et al.\ (\cite{zane02}), Kaplan et al.\ 
(\cite{kaplanetal02}) and Cropper et al.\ (\cite{cropper})
from X-ray measurements of the neutron-star spindown rate.
We adopt the central value $t=1300$ kyr with an uncertainty
by a factor of 2.
The ages of three other pulsars, PSR B1706--44, Geminga,
and PSR B1055--52, are  the
characteristic pulsar ages assuming 
an uncertainty by a factor of 2.

For two youngest sources, RX J0205+6449 and the Crab pulsar, no 
thermal emission has been detected, but the upper limits on 
surface temperature $T_{\rm s}^\infty$ have been established
(Slane et al.\ \cite{slane02}, Weisskopf et al.\ \cite{weisskopf04}).
The surface temperatures of the next five sources,
RX J0822--4300, 1E 1207.4--5209,
Vela, PSR B1706--44, and PSR J0538+2817,
have been obtained using hydrogen atmosphere models
(see references in Table 1).
Such models are more consistent with other information 
on these sources (e.g., Pavlov et al.\ \cite{pavlovetal02}) 
than the blackbody
model. On the contrary, for the Geminga and PSR B1055--52
we present the values
of $T_{\rm s}^\infty$ inferred using the blackbody
spectrum because this spectrum is more consistent for these sources.

Let us notice that from Table 1 we have excluded PSR B0656+14
which was considered earlier (e.g., Yakovlev et al.\ \cite{yakovlevetal02}).
A combined analysis of new X-ray and optical observations
of the source (with the improved distance from
new parallax measurements of Brisken et al.\ \cite{briskenetal03})
leads either to unrealistically small values of
the neutron star radius (in the blackbody model)
or to unreasonably small distance to the star (in the hydrogen
atmosphere model); see, e.g., Zavlin \& Pavlov (\cite{zp02}).
This makes current interpretations of the data unreliable.

The surface temperature of RX J1856.4--3754 is still rather
uncertain. A wide scatter of $T_{\rm s}^\infty$,
obtained by different authors,
takes place because
X-ray and optical observations are not described
by one blackbody model. This can be explained, for instance,
by the presence of hot spots on the neutron star
surface. Thus, we adopt the upper limit $T_{\rm s}^\infty
< 0.65$ MK, which agrees with the value of $T_{\rm s}^\infty$
obtained either with the ``Si-ash'' atmosphere model
of Pons et al.\ (\cite{ponsetal02}) or with the model of condensed surface
layers of Burwitz et al.\ (\cite{burwitzetal03}). It agrees also with the
model of nonuniform surface temperature distribution
suggested by Pavlov \& Zavlin (\cite{pz03}). In the latter case,
the mean surface temperature $T_{\rm s}^\infty \approx 0.5$ MK
is below our upper limit of $T_{\rm s}^\infty$.

Finally, $T_{\rm s}^\infty$ for RX J0720.4--3125
is taken from Motch et al.\ (\cite{motchetal03})
who have interpreted the observed spectrum
with a model of a hydrogen atmosphere of finite depth.

For PSR J0538--4300, PSR B1055-52, and RX J0720.4--3125,
the authors cited in Table 1
have not reported any error bars of $T_{\rm s}^\infty$.
We adopt 20\% uncertainties
which seem to be appropriate for these sources.

%%%%%%%%%%%%%%%%%%%%%%%%%%%%%%%%%%%%%%%%%%%%%%%%%%%%%%%%%%%%%%%%%%%%%%%%%%%
\section{Physics input and calculations}
\label{physics}
%%%%%%%%%%%%%%%%%%%%%%%%%%%%%%%%%%%%%%%%%%%%%%%%%%%%%%%%%%%%%%%%%%%%%%%%%%%

\begin{figure*}[t]
%\begin{center}
\centering
\epsfysize=80mm
\epsffile[18 145 569 418]{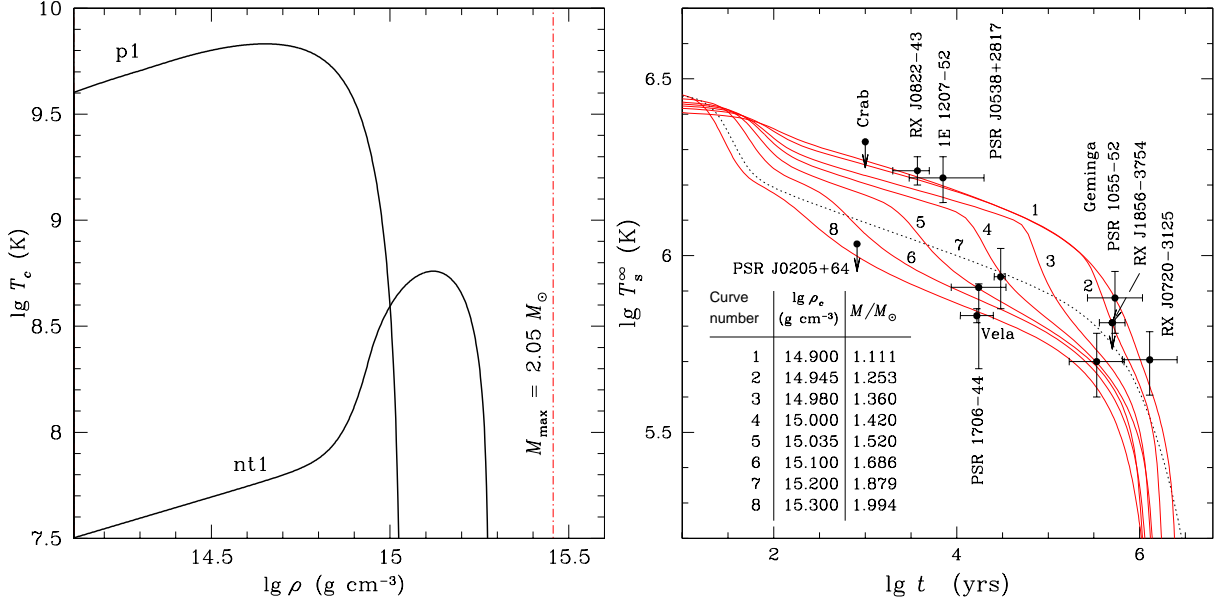}
\caption{{\it Left:} Density dependence of
critical temperature of
model p1 for proton superfluidity and model nt1
for neutron superfluidity in a neutron-star core;
vertical dot-and-dash line indicates the
central density of a maximum-mass neutron star.
{\it Right:} Observations (Table 1) compared
with theoretical cooling curves of eight
neutron stars (1--8) with different masses.
All solid curves refer to neutron stars
with model superfluidities from the left panel.
The dotted curve 7 is for a non-superfluid star.
Insert table gives masses and central densities
of stars 1--8.
}
%\end{center}
\label{cool}
\end{figure*}

We will simulate cooling of neutron stars using our
generally relativistic cooling code described by
Gnedin et al.\ (\cite{gyp01}). 
We adopt a moderately stiff equation of
state of neutron star interiors proposed by
Douchin \& Haensel (\cite{dh01}). According to this equation of state,
neutron star cores (regions of the densities
$\rho > 1.3 \times 10^{14}$ g~cm$^{-3}$) 
consist of neutrons, with the admixture
of protons, electrons and muons. All constituents
exist everywhere in a core, except for muons
which appear at $\rho> 2.03 \times 10^{14}$
g~cm$^{-3}$. 
The most massive stable
star has the (gravitational) mass $M=M_{\rm max}=2.05\,M_\odot$,
the central density $\rho_{\rm c}=2.9 \times 10^{15}$
g~cm$^{-3}$, and the (circumferential) radius $R=9.99$ km.  
The central densities and masses of eight neutron
star models (with $M$ from 1.111~$M_\odot$ to 1.994~$M_\odot$)
are presented in the right panel of Fig.\ \ref{cool}.

All physics input is standard. The effects of muons
are included as described by Bejger et al.\ (\cite{byg03}).
We assume no envelope of light elements on stellar surfaces
(Sect.\ \ref{testing}).
The code calculates the cooling curves, which give the dependence
of the effective surface stellar temperature $T_{\rm s}^\infty$
on stellar age $t$.
Let us remind that neutron stars are born hot in supernova explosions
(with internal temperatures $T \sim 10^{11}$ K) but gradually
cool down via neutrino emission from the entire
stellar body and via heat diffusion to the surface
and thermal surface emission
of photons. Qualitatively, one can distinguish three cooling 
stages. At the first 
(`non-isothermal') stage ($t \la 100$ yr)
the main cooling mechanism is neutrino emission but
the stellar interior stays highly non-isothermal.
At the second (`neutrino') stage ($10^2 \la t \la 10^5$ yr) the
cooling goes mainly via neutrino emission from isothermal
interiors.  At the third (`photon') stage ($t \ga 10^5$ yr) 
a star cools predominantly through the surface photon emission.

%%%%%%%%%%%%%%%%%%%%%%%%%%%%%%%%%%%%%%%%%%%%%%%%%%%%%
%%%%%%%%%%%%%%%%%%%%%%%%%%%%%%

The new element of our present studies is the equation of state of 
Douchin \& Haensel (\cite{dh01}). We have chosen it
because it forbids the powerful direct Urca
process of neutrino emission (Lattimer et al.\ \cite{lpph91})
in all stable neutron
stars ($M \leq M_{\max}$). In this case, a non-superfluid neutron
star of any mass $M_\odot \la M \leq M_{\rm max}$
will have almost the same (universal) cooling curve
$T_{\rm s}^\infty(t)$ (the dotted curve
in the right panel of Fig.\ \ref{cool}).
At the neutrino cooling stage, this curve is determined
by the neutrino emission due to the modified Urca process.
The curve is almost independent
of the equation of state of neutron star cores 
(Page \& Applegate \cite{pa92}) as long
as the direct Urca process is forbidden. 
As has been indicated by many authors
(see, e.g., Yakovlev \& Pethick \cite{yp04} and references therein) and
seen from Fig.\ \ref{cool},
this universal cooling model is certainly
unable to explain the data. For instance, it gives
$T_{\rm s}^\infty$ much lower than that of PSR B1055--52,
but much higher than that of the Vela pulsar.
We will show that all the data can be explained
assuming superfluidity of neutron-star cores.

It is well known that neutrons and protons in stellar cores
can be in superfluid state.
Proton superfluidity is caused by singlet-state proton
pairing, while neutron superfluidity is produced
by triplet-state neutron pairing. These superfluidities
can be specified by density dependent critical temperatures
for protons and neutrons,
$T_{\rm cp}(\rho)$ and $T_{\rm cn}(\rho)$. 
Results of calculations of these temperatures
from microscopic theories show
a large scatter of critical temperatures depending on
a nucleon-nucleon interaction model and a many-body theory
employed. In particular, recently Schwenk \& Friman
(\cite{sf04}) and Zuo et al.\ (\cite{zuo}) have obtained
weak neutron and proton pairing in neutron star cores
but many other calculations give much stronger superfluidity 
(e.g., Lombardo \& Schulze \cite{ls01};
also see references in Yakovlev et al.\ \cite{yls99}).
In this situation it is reasonable to
consider $T_{\rm cp}(\rho)$ and $T_{\rm cn}(\rho)$
as unknown functions of $\rho$ (consistent with
predictions of microscopic theories) which can
hopefully be constrained 
by comparing theoretical cooling
curves with observations.

Superfluidity of neutrons and/or protons in neutron-star
cores affects the heat capacity of nucleons and reduces
neutrino reactions (Urca and nucleon-nucleon
bremsstrahlung processes) involving superfluid nucleons
(as reviewed, e.g., by Yakovlev et al.\ \cite{yls99}). Moreover,
superfluidity initiates an additional neutrino emission
mechanism associated with Cooper pairing of 
nucleons (Flowers et al.\ \cite{frs76}). All these effects
of superfluidity are incorporated into our cooling code.

In our calculations we adopt one model of strong
superfluidity of protons (with the maximum of $T_{\rm cp}(\rho)$
about $T_{\rm cp}^{\rm max} \approx 7 \times 10^9$ K) 
and several models of moderate
superfluidity of neutrons (with 
$T_{\rm cn}^{\rm max} \sim 6 \times 10^8$ K) in a neutron-star core.
These models are phenomenological but consistent
with the results of microscopic theories.
A pair of models: proton superfluidity p1 and
neutron superfluidty nt1 is plotted in the left panel
of Fig.\ \ref{cool}.

The strong proton superfluidity is required
to slow down the cooling of low-mass stars, $M \la 1.1\,M_\odot$,
whose central densities are 
$\rho_{\rm c} \la 8 \times 10^{14}$ g cm$^{-3}$. 
This scenario was suggested by
Kaminker et al.\ (\cite{khy01}). In a low-mass star,
one has $T_{\rm c}(\rho) \ga 3 \times 10^9$~K
everywhere in the core.
The proton superfluidity occurs at the early
cooling stage ($t \la 1$ yr) and suppresses
modified Urca processes of neutrino emission
as well as neutrino generation in proton-proton and
proton-neutron collisions. Neutrino emission due to
Cooper pairing of protons is switched on too early
and becomes inefficient in middle-aged neutron stars
we are interested in. 
In contrast, the adopted neutron superfluidity is too weak
in low-mass stars (the left panel of Fig.\ \ref{cool})
to appear at the neutrino cooling stage.
This superfluidity does not suppress
the neutrino emission in neutron-neutron collisions
which becomes the leading mechanism
of neutrino cooling. It is much weaker than
the modified Urca process (which would be leading in non-superfluid
stars). As a consequence, the cooling curves
of low-mass stars go noticeably higher than the universal
cooling curve of non-superfluid stars.
Actually, these cooling curves also merge
into one almost universal curve,
which is independent of the equation of state in a stellar
core and the exact behaviour of the $T_{\rm cp}(\rho)$
(Kaminker et al.\ \cite{kyg02}). This upper curve 1 allows one to explain
observations of neutron stars hottest for their age
(RX J0822--4300, 1E 1207.4--5209, PSR B1055--52, RX J0720.4--3125)
as cooling low-mass neutron stars.

Now we come to observations of neutron stars coldest
for their age (first of all, PSR J0205+6449, the Vela pulsar,
and Geminga). It has been widely proposed to
interpret these objects as rather massive neutron stars
with the neutrino emission enhanced by direct Urca process
in nucleon cores (or by similar processes in 
pion-condensed, kaon-condensed or quark cores).
We will show that coldest objects can be explained without
invoking these mechanisms by tuning the model of moderate neutron
superfluidity at $\rho \ga 8 \times 10^{14}$ g cm$^{-3}$.
Let us consider the most massive neutron star ($1.994\,M_\odot$,
curve 8) in Fig.\ \ref{cool}. Its central density is higher
than the density at which neutron superfluidity nt1 dies out.
When the internal temperature of the star becomes lower
than the maximum critical temperature of the neutron superfluidity,
the neutrino emission due to Cooper pairing of neutrons
switches on and becomes a powerful neutrino emission mechanism,
which can be about two orders of magnitude more
efficient than the modified Urca process in a
non-superfluid star (see Sect.\ \ref{CP}). This emission
produces {\it enhanced} cooling (attributed to direct
Urca or similar processes in previous calculations).
The enhancement is not too strong (e.g., the direct Urca
process in a nucleon 
stellar core would further enhance the neutrino luminosity
by about 4--5 orders of magnitude). However, even this not too strong
enhancement is sufficient to explain observations
of the coldest neutron
stars (particularly, PSR J0205+6449, the Vela and Geminga pulsars).
Evidently, all neutron stars with $\rho_{\rm c} \ga 2 \times 10^{15}$
g~cm$^{-3}$ (in our model) will cool nearly as fast as the 
$1.994\,M_\odot$ star
in Fig.\ \ref{cool}.

Therefore, we come to three distinct classes of cooling neutron stars
(similar to those described by Kaminker et al.\ \cite{kyg02} for the case
of enhanced cooling due to direct Urca process). The first
class contains low-mass, very slowly cooling stars (curve 1
in the right panel of Fig.\ \ref{cool}). Another class contains
high-mass stars with enhanced cooling (curve 8). Finally, there
is a class of medium-mass neutron stars (curves 2--6)
which show intermediate cooling. Their cooling curves
fill in the space between the upper curve for low-mass
stars and the lower curve for high-mass stars. These curves
explain observations of PSR B1706--44, PSR J0538+2817,
and RX J1856.4--3754.   

%%%%%%%%%%%%%%%%%%%%%%%%%%%%%%%%%%%%%%%%%%%%%%%%%%%%%%%%%%%%%%%%%%%%%%%%%%%
\section{Cooper-pairing neutrino emission as a fast-cooling agent}
\label{CP}
%%%%%%%%%%%%%%%%%%%%%%%%%%%%%%%%%%%%%%%%%%%%%%%%%%%%%%%%%%%%%%%%%%%%%%%%%%%

Let us give a simple explanation of the computer results
on enhanced neutrino emission due to Cooper pairing of neutrons.
We start from the expression
for the neutrino emissivity $Q_{\rm CP}$ due this
process (e.g., Eq.\ (236) in Yakovlev et al.\ \cite{ykhg01}). 
It can be written as
\begin{equation}
    Q_{\rm CP}(\rho,T)=q(\rho,T)\,F(\tau),
\label{cooling-QCP}
\end{equation}
where 
\begin{eqnarray}
  q(\rho,T) & \approx &
  1.17 \times 10^{21} \, \left( { m_N^\ast \over m_N } \right) \,
     \left( { p_{\rm F} \over m_N c } \right) 
\nonumber \\     
     & & \times T_9^7 \, {\cal N}_\nu \;
     a_N \; \; {\rm erg \;cm^{-3} \; s^{-1}},
\label{cooling-q(rhoT)}
\end{eqnarray}
$T \equiv T_9 \times 10^9$ K is the internal stellar temperature,
$m_N$ is the bare nucleon ($N=n$ or $p$) mass,
$m_N^\ast$ is the nucleon effective mass in dense matter,
$p_{\rm F}$ is the nucleon Fermi momentum,
$a_N$ is a dimensionless constant combined of
squared weak-interaction constants of vector
and axial-vector nucleon currents,
${\cal N}_\nu=3$ is the number of neutrino flavors,
and $F(\tau)$ is a function of $\tau=T/T_{\rm c}$.
The constant $a_N$ depends on nucleon species and pairing type,
while $F(\tau)$ depends on pairing type.
We have $a_n=4.17$ for the triplet-state neutron
pairing under discussion. This value can be renormalized
by many-body effects (for instance,
the renormalization of the axial-vector constant
was considered by Carter \& Prakash
\cite{cp02}). However, theoretical cooling curves
are not too sensitive to the exact value of $a_n$,
and we use the non-renormalized value. The analytic
fit expression for $F(\tau)$ is presented, for instance,
by Yakovlev et al.\ (\cite{ykhg01}). Let us remind that
$F(\tau) \approx 4.71\,(1-\tau)$ just after superfluidity
onset (immediately after $T$ falls below $T_{\rm c}$) and
$F(\tau) \approx 1.27\, \tau^{-6}\, \exp(-2.376/\tau)$
at $\tau \ll 1$. Thus, the emissivity $Q_{\rm CP}(\rho,T)$
is exponentially suppressed at $T \ll T_{\rm c}$.

%%%%%%%%%%%%%%%%%%%%%%%%%%%%%%%%%%%%%%%%%%%%%%%%%%%%%%%%%%%%%%
\begin{figure}
\centering
\epsfxsize=\hsize
\epsffile[85 220 550 680]{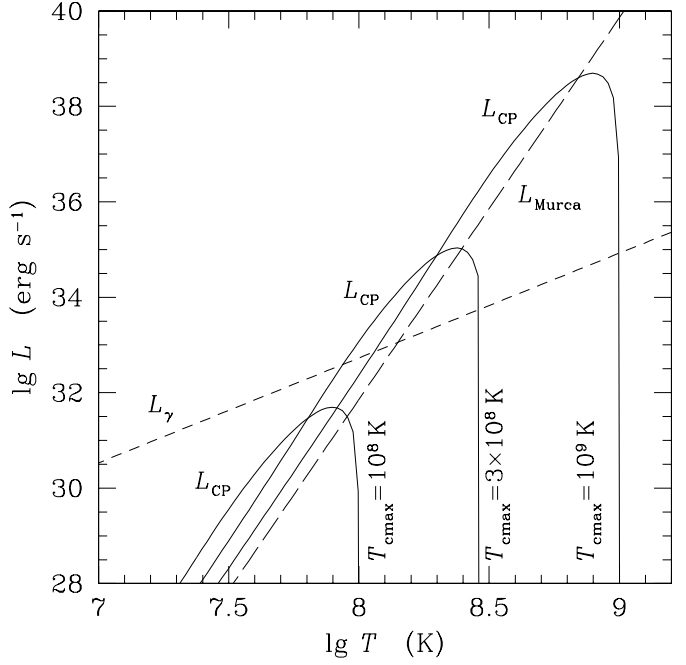}
\caption{ A sketch of neutrino luminosities produced
by the modified Urca process ($L_{\rm Murca}$) 
and Cooper pairing process ($L_{\rm CP}$)
as well as of the photon luminosity $L_\gamma$ 
of a neutron star versus internal temperature
$T$ for three models of neutron superfluidity in the stellar
core with $T_{\rm cn}^{\rm max}=10^8$, $3 \times 10^8$ and
$10^9$~K.
}
\label{fig-cooling-cpluma}
\end{figure}
%%%%%%%%%%%%%%%%%%%%%%%%%%%%%%%%%%%%%%%%%%%%%%%%%%%%%%%%%%%%%%%

For our qualitative analysis in this section we 
employ  the simplest dependence of the neutron
critical temperature on distance $r$ from the stellar center:
\begin{equation}
     T_{\rm cn}(r)= T_{\rm cm} \left\{ 1 
     - {(r-r_{\rm m})^2 \over (\Delta r_{\rm m})^2} \right\}
\label{cooling-tc}
\end{equation}
at $|r-r_{\rm m}| < \Delta r_{\rm m}$ (with the maximum
$T_{\rm cm}=T_{\rm cn}^{\rm max}$ at $r=r_{\rm m}$), and $T_{\rm cn}=0$
at $|r-r_{\rm m}| \geq \Delta r_{\rm m}$. 

Neglecting, for simplicity, general relativistic effects and
assuming an isothermal stellar core at a temperature $T<T_{\rm cm}$,
the neutrino luminosity $L_{\rm CP}$ due to
Cooper pairing of neutrons can be written as
\begin{equation}
      L_{\rm CP}= 4 \pi \int_{r_1}^{r_2} r^2\, Q_{\rm CP}\,
                   {\rm d}r.
\label{cooling-LCP}
\end{equation}
Here, $r_1$ and $r_2$ restrict the superfluid layer, where $T<T_{\rm cn}$
and the neutrino process in question is allowed. To be specific,
let us assume that the widest superfluid layer 
(which is realized at $T=0$ and extends
from $r_{\rm m}-\Delta r_{\rm m}$ to $r_{\rm m}+\Delta r_{\rm m}$)
entirely falls in the neutron star core.
  
The factor $F(\tau)$ in the emissivity $Q_{\rm CP}$,
Eq.\ (\ref{cooling-QCP}), is a more rapidly varying function of $r$
than $q(\rho,T)$. Thus we can set $r=r_{\rm m}$ and
$q(\rho,T)=q(\rho_{\rm m},T)$ 
(with $\rho_{\rm m}=\rho(r_{\rm m})$) in all functions 
under the integral but in
$F(\tau)$. A simple replacement
of integration variable leads then to
\begin{eqnarray}
      L_{\rm CP}&= &8 \pi r_{\rm m}^2 \, \Delta r_{\rm m} \,
         q(\rho_{\rm m},T)\, \tau_{\rm m} \, \ell (\tau_{\rm m}),
\label{cooling-LCP1} \\
      \ell (\tau)& = & {1 \over 2} \int_\tau^1 {{\rm d}\tau' \, F(\tau')
        \over \tau^{\prime\, 3/2} \sqrt{\tau'-\tau}},
\label{cooling-LCP2}
\end{eqnarray}
where $\tau_{\rm m}=T/T_{\rm cm}$.
The integration can be done numerically; the appropriate
analytic fit (for triplet-state neutron pairing) is
\begin{eqnarray}
  {\ell}(\tau)& \!= \!\!&
    (1-\tau)^{3/2} \,
   \Big[ 3.844 \,(1-\tau)+3.142\,\tau^2 \Big.
\nonumber \\
     & \! + \!\!&
        13.99\tau(1-\tau) + \left.
        {25.4 \, \tau^{2.5}\,(1-\tau)^2
        \over ((\tau-0.2493)^2+0.03694)^{0.7}}\right].
\label{cooling-l(tau)}
\end{eqnarray}
%
%We can also rewrite Eq.\ (\ref{cooling-LCP1})
%as $L_{\rm CP}=8 \pi r_{\rm m}^2 \, \Delta r_{\rm m} \,
%q(\rho_{\rm m},T_{\rm cm})\, L_{\rm nm}(\tau_{\rm m})$,
%where the dimensionless function $L_{\rm nm}(\tau)=\tau^8 {\ell}(\tau)$ 
%is the normalized neutrino luminosity which describes
%the temperature dependence of $L_{\rm CP}$. 

Evidently, the luminosity $L_{\rm CP}$ vanishes in a hot star, 
where $T>T_{\rm cm}$ and neutron superfluidity is absent.
It switches on as $T$ falls below $T_{\rm cm}$;
it grows almost linearly while $T$ decreases to $\sim 0.8\, T_{\rm cm}$;
afterwards, it
reaches maximum at $T=0.792\,T_{\rm cm}$ 
(with $\tau\, {\ell}(\tau)=0.792 \, {\ell}(0.792)
%
% {\ell}(0.792)=0.6077
%
=0.481$)
and then decreases.
At the increasing and maximum-luminosity stage, 
$L_{\rm CP}$ is collected from a superfluid spherical
stellar layer in the vicinity of the maximum critical temperature,
$r \approx r_{\rm m}$.
This creates a splash of neutrino emission associated
with Cooper pairing of neutrons. 

For typical values of the parameters, 
the maximum value of $L_{\rm CP}$ can be 
one-two orders of magnitude higher than the neutrino
luminosity $L_{\rm Murca}$ of a non-superfluid 
star (with forbidden direct Urca process). This is demonstrated in
Fig.\ \ref{fig-cooling-cpluma}
using a toy model of cooling neutron stars
described by Yakovlev \& Haensel (\cite{yh03}) -- there is no need
to employ accurate models in this section.
The parameters of the neutron-star model presented at the figure
are: $M=1.16\, M_\odot$, $R=12$ km, $\rho_{\rm c}=8 \times 10^{14}$
g cm$^{-3}$, $r_{\rm m}=\Delta r_{\rm m}=5$ km.
The three superfluidity models ($T_{\rm cn}(r)$) are selfsimilar and
differ by the values of $T_{\rm cm}=10^8$, $3 \times 10^8$,
and $10^9$ K.   
Three solid lines exhibit the Cooper-pairing
neutrino luminosity $L_{\rm CP}$ calculated from Eqs.\ 
(\ref{cooling-LCP})--(\ref{cooling-l(tau)}) for
three models of neutron superfluidity.
% presented
%with different $T_{\rm cm}$.
Since $L_{\rm Murca} \propto T^8$
and $L_{\rm CP}^{\rm max} \propto T^7$, the Cooper-pairing
luminosity is more competitive at weaker superfluidity
(lower $T_{\rm cm}$). 
However, at $T_{\rm cm} \la 2 \times 10^8$ K this luminosity
becomes lower than the photon thermal luminosity of the star
(Fig.\ \ref{fig-cooling-cpluma}) 
which makes it insignificant for stellar cooling.
It is worth to notice that, for
realistic parameters, $L_{\rm CP}$ is much smaller
than the neutrino luminosity due to the direct Urca process in 
a non-superfluid star (if the direct Urca process is open).
   
The decreasing part of $L_{\rm CP}(T)$ is even more fascinating.
We have ${\ell}(\tau)\approx 3.84$ as $\tau \to 0$, resulting
in the scaling relation
\begin{equation}
L_{\rm CP} \propto \Delta r_{\rm m} T^8/T_{\rm cm},
\label{scaling}
\end{equation}
which becomes sufficiently accurate at $T \la 0.6\,T_{\rm cm}$.
This neutrino emission is actually produced from two
thin spherical shells (near $r=r_1$ and $r=r_2$), 
where $T$ is just below $T_{\rm c}(r)$. 
The widths of these shells are proportional to $T$,
which explains the power-law $T^8$ (instead of
the exponential decrease of the emissivity $Q_{\rm CP}(\rho,T)$
in a local element of superfluid matter). Therefore,
the decreasing part of the Cooper-pairing neutrino luminosity
has the {\it same temperature dependence} as 
all slow
neutrino emission mechanisms (modified Urca, nucleon-nucleon
bremsstrahlung) in non-superfluid cores of stars
with forbidden direct Urca processes.
In other words, the superfluidity suppresses the neutrino
emission available in non-superfluid stars but  
initiates the Cooper-pairing neutrino emission 
in such a way that it 
{\it acts as a new nonsuppressed  
neutrino cooling mechanism}. Moreover,
the new emission can be more
intense than that in a non-superfluid star 
and provide enhanced cooling. 
This important feature
appears in realistic models of cooling
neutron stars with density dependent critical
temperatures $T_{\rm c}(\rho)$ (and does not appear
in the models with density-independent $T_{\rm c}$).
In particular, it implies that once $L_{\rm CP}$
takes on leadership in competition with $L_{\rm Murca}$
just after the superfluidity onset, it will not lose it
during subsequent evolution (especially because
$L_{\rm Murca}$ is actually suppressed by superfluidity,
which is not taken into account in Fig.\ \ref{fig-cooling-cpluma}).
This is clearly seen from Fig.\ 
\ref{fig-cooling-cpluma}.

Let us add that at $T \ll T_{\rm cm}$ we
can obtain a better formula for $L_{\rm CP}$ than
Eq.\ (\ref{cooling-LCP}), without employing the specific
$T_{\rm c}(r)$ profile, Eq.\ (\ref{cooling-tc}).
It is sufficient to start from Eq.\ (\ref{cooling-LCP})
and notice that the main contribution into
$L_{\rm CP}$ comes from two thin shells, at $r\approx r_1$
and $r \approx r_2$, where $T_{\rm c}(r) \approx T$.
In each shell, the gradient $D={\rm d}T_{\rm c}(r)/{\rm d}r$
can be taken constant. Then we get
\begin{equation}
      L_{\rm CP}= 8 \pi \left[ 
         r_1^2 \, H_1 \, q(\rho_1,T)+     
         r_2^2 \, H_2 \, q(\rho_2,T) \right] \ell (0),
\label{cooling-LCP3}
\end{equation}
where $H_1=T/|D_1|$ and $H_2=T/|D_2|$ are characteristic
widths of our shells, $\rho_1=\rho(r_1)$, $\rho_2=\rho(r_2)$,
and $\ell(0)=3.84$.
Strictly speaking, $r_1$, $r_2$, $\rho_1$, $\rho_2$,
$D_1$, and $D_2$ depend slightly on
$T$, but this dependence can be regarded as parametric.
It is easy to verify that if $T_{\rm c}(r)$ is
given by Eq.\ (\ref{cooling-tc}) at $T \ll T_{\rm cm}$
and $\Delta r_{\rm} \ll r_{\rm m}$,
our new expression for $L_{\rm CP}$ coincides with
Eq.\ (\ref{cooling-LCP1}). 
Equation (\ref{cooling-LCP1}) is expected to be
useful just after the superfluidity onset, 
at $0.6 \,T_{\rm cm} \la T < T_{\rm cm}$
(where the parabolic $T_{\rm c}(r)$ dependence may be a good
approximation),
while Eq.\ (\ref{cooling-LCP3}) is more exact at lower $T$.   
Both equations enable one to incorporate the Cooper-pairing 
neutrino emission in simplified cooling models
(like a toy model of Yakovlev \& Haensel \cite{yh03}),
useful for understanding main features of neutron star
cooling without complicated cooling codes.

The above analysis is valid as long as $T_{\rm c}(\rho)$
vanishes in the stellar interior. If not, there is
a minimum value $T_{\rm c}^{\rm min}$ of $T_{\rm c}(\rho)$,
and $L_{\rm CP}$ will become exponentially suppressed at
$T \ll T_{\rm c}^{\rm min}$.

%%%%%%%%%%%%%%%%%%%%%%%%%%%%%%%%%%%%%%%%%%%%%%%%%%%%%%%%%%%%%%%%%%%%%%%%%%%%%%%%%%%%%%%%
\section{Testing the cooling scenario and discussion}
\label{testing}
%%%%%%%%%%%%%%%%%%%%%%%%%%%%%%%%%%%%%%%%%%%%%%%%%%%%%%%%%%%%%%%%%%%%%%%%%%%%%%%%%%%%%%%%

After clarifying the efficiency of the 
Cooper-pairing neutrino emission let us return to
the cooling scenario described in Sect.\ \ref{physics}.
As we have already mentioned, 
the scenario is rather insensitive to a specific
model of proton superfluidity (required to raise the
surface temperature of low-mass stars for explaining 
observations of the sources hottest for their ages).
The only serious constraint on the proton pairing is
that $T_{\rm cp}(\rho)$ should be high ($\ga 3 \times 10^9$ K)
in the cores of low-mass stars.

%%%%%%%%%%%%%%%%%%%%%%%%%%%%%%%%%%%%%%%%%%%%%%%%%%%%%
\begin{figure*}[t]
%\begin{center}
\centering
\epsfysize=80mm
\epsffile[18 145 569 418]{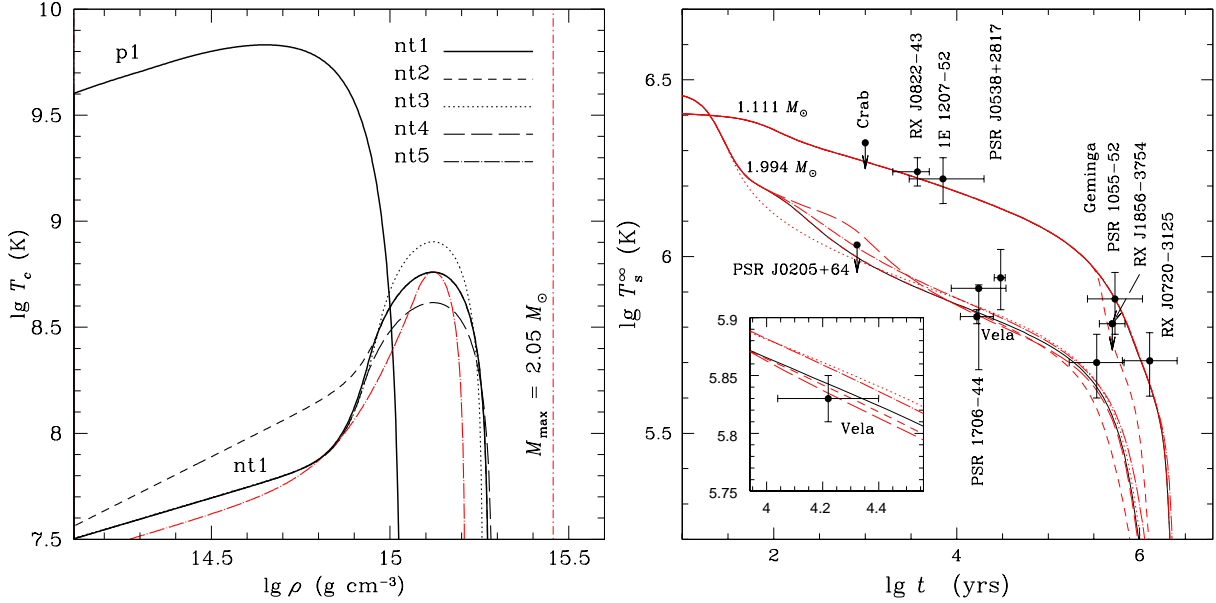}
\caption{{\it Left:} One model p1 for proton superfluidity
and five models nt1--nt5 for neutron superfluidity
in a neutron-star core. {\it Right:} Cooling curves
of low-mass (1.111 $M_\odot$) and
high-mass (1.994 $M_\odot$) stars with model p1 proton superfluidity
and one of the models of neutron superfluidity from the left
panel, compared with observations. Cooling of the low-mass
star is insensitive to selected models of neutron superfluidity
(except for model nt2  at $t > 300$ kyr).
Insert shows the comparison of cooling curves of the high-mass star
with observations of the Vela pulsar in more details. }
%\end{center}
\label{test}
\end{figure*}
%%%%%%%%%%%%%%%%%%%%%%%%%%%%%%
%

However, the constraints on the neutron critical 
temperature $T_{\rm cn}(\rho)$ in a stellar core
{\it should be really strong}. This is illustrated in Fig.\ \ref{test}.  
The left panel  displays the critical temperatures
of our proton superfluidity model (p1) and five neutron superfluidity
models (nt1--nt5), including our basic model nt1 used
in Sect.\ \ref{physics}. The right panel shows cooling curves
of a low-mass ($1.111\,M_\odot$) star and
a high-mass ($1.994\,M_\odot$) star.
Any curve is calculated for model p1 of the proton superfluidity
and one model of the neutron superfluidity from the left
panel of Fig.\ \ref{test}. 
Any observational point between an upper curve
and a lower curve can be explained
by a given superfluid model.
The constraints 
on the neutron superfluidity are as follows.

({\it 1}) The neutron superfluidity should be weak
in low-mass stars. In our case (for the equation of state
of Douchin \& Haensel \cite{dh01}) this means that
$T_{\rm cn}(\rho) \la 2 \times 10^8$ K at $\rho \la 8 \times 10^{14}$
g~cm$^{-3}$. Under this condition the neutron superfluidity does
not affect the cooling 
(at least at the neutrino cooling stage)
of low-mass stars ($M \la 1.1\,M_\odot$) 
and does not violate our interpretation of
the sources hottest for their age (first of all, 
RX J0822--4300 and PSR B1055--52).
Accordingly, all five cooling curves (for superfluids nt1--nt5)
of low-mass stars merge in one upper (solid) cooling curve in
Fig.\ \ref{test}. The only exclusion is provided by
model nt2 with highest pre-peak $T_{\rm cn}(\rho)$
among models nt1--nt5. In a low-mass star this superfluidity occurs at
$t \ga 300$ kyr. The Cooper-pairing
neutrino emission and reduced heat capacity of neutrons
noticeably accelerate the cooling at this late stage
(the upper short-dashed curve).

({\it 2}) The neutron superfluidity should be moderate
at $\rho \ga 10^{15}$ g~cm$^{-3}$, with the peak maximum
$T_{\rm cn}^{\rm max} \sim 6 \times 10^8$ K (model nt1 
in Fig.\ \ref{test}, the solid 
curve). In this case it switches on
just in time to initiate
the enhanced cooling in a high-mass star. 
Its level is sufficient to explain
observations of neutron stars coldest for their ages
(first of all, PSR J0205+6449 and the Vela pulsar).
The asymptotic neutrino-cooling regime given by the scaling expression
(\ref{scaling}) is realized at $t \ga (1-10)$ kyr.
If $T_{\rm cn}^{\rm max}$ were slightly higher
than $6 \times 10^8$ K (model nt3, 
$T_{\rm cn}^{\rm max}=8 \times 10^8$ K, the dotted curve),
the Cooper-pairing neutrino emission will start
operating in a younger massive star but becomes less
efficient at $t \sim 10$ kyr, which is less favorable for
explaining the observations of the Vela pulsar.
This cooling behaviour is naturally explained by the scaling (\ref{scaling}).
If $T_{\rm cn}^{\rm max}$ were slightly lower
than $6 \times 10^8$ K (model nt4, 
$T_{\rm cn}^{\rm max}=4 \times 10^8$ K,
the long-dashed
curve), the Cooper-pairing neutrino emission will start
operating too late which would violate the interpretation
of the observations of PSR J0205+6449.  

({\it 3}) The results are also sensitive to the width
of the peak of the $T_{\rm cn}(\rho)$ curve.
For instance,  retaining the peak maximum of $6 \times 10^8$ K
but making the peak narrower (model nt5, the dot-dashed
curve) will reduce the neutrino emissivity due
to neutron pairing, raise the temperature
of the massive star and complicate the interpretation
of the Vela pulsar
(again, in agreement with the scaling
(\ref{scaling})). However, the cooling curves
are rather insensitive to the exact position of
the $T_{\rm cn}(\rho)$ maximum.
We can slightly shift the maximum to higher
or lower $\rho$ (confining the peak within the
kernel of a massive star) but these shifts will
not change the cooling curves of massive stars
(such tests are not shown in Fig.\ \ref{test}). 
However, the shift of the maximum to $\rho \la 8 \times
10^{14}$ g cm$^{-3}$ would cause the enhanced
cooling of low-mass stars.  
The cooling curves of low-mass stars would become
close to those of high-mass stars which 
would violate the interpretation
of neutron stars hottest for their ages (see item ({\it 1})).
 
This discussion shows that the cooling curve of a
massive neutron star implying model nt1 of neutron
superfluidity is close to the {\it lowest cooling curve}
(in the scenario, where the cooling
is enhanced by Cooper-pairing neutrino emission).
Observations of cold neutron stars,
PSR J0205+6449 and the Vela pulsar,
provide excellent tests for this scenario.  
If these pulsars were
noticeably colder we would be unable to explain them
within our scheme. Notice, that the upper limit of the
surface temperature of PSR J0205+6449 was inferred from observations
(Slane et al.\ \cite{slane02}) using the blackbody spectrum of surface
emission. If this pulsar has a hydrogen atmosphere,
the upper limit on $T_{\rm s}^\infty$ could be expected to be
about twice lower than for the blackbody case. In that case
we would be unable to explain this source within the proposed
scenario.

Although we have used one equation of state of dense
matter (Douchin \& Haensel \cite{dh01}) we would obtain similar
results for other equations of state which forbid direct
Urca processes (and other similar processes of fast
neutrino cooling) in neutron star cores. Taking different
equations of state would lead to attributing different masses
to the same sources (Fig.\ \ref{cool}); similar
problem has been discussed by Kaminker et al.\ (\cite{kyg02}).

%As discussed above (item ({\it 1})), we need weak neutron
%pairing $T_{\rm cn}(\rho) \la 2 \times 10^8$ K in 
%low-mass stars (at $\rho \la 8 \times 10^{14}$ g cm$^{-3}$,
%$M \la 1.1\,M_\odot$, 
%for the equation of state of Douchin \& Haensel \cite{dh01}).

In addition, we could take an equation of state
in the stellar core which opens direct Urca
process at highest densities (in the central kernels
of most massive stable neutron stars; similar to the
equation of state of Akmal \& Pandharipande \cite{ap98}).
Applying the same model of nucleon superfluidty
as in Fig.\ \ref{cool},
we would get {\it five} types of cooling neutron stars
(instead of three). Three types would be the same as those
mentioned in Sect.\ \ref{physics}: low-mass, 
very slowly cooling 
stars; massive stars whose cooling is enhanced
by Cooper-pairing neutrino emission; and 
of medium-mass stars whose cooling is intermediate.
In addition, we would have: most massive neutron stars
demonstrating a very fast cooling via the direct
Urca process; and 
stars whose cooling is intermediate between that
enhanced by the Cooper-pairing neutrino emission and
by the direct Urca process. The transition from 
the Cooper-pairing neutrino cooling to direct-Urca cooling
with increasing mass $M$ will be very sharp and
the number of intermediate-cooling sources 
will be small.
The maximum-mass neutron
stars would be extremely cold
($T_{\rm s}^\infty \sim 2 \times 10^5$ K
at $t \sim 10$ kyr), about the same as discussed, e.g., by Kaminker
et al.\ (\cite{kyg02}). A discovery of such stars would definitely
indicate the operation of the direct Urca process in their
cores. An indirect evidence of their existence is
provided by a non-detection of neutron stars in
some supernova remnants (Kaplan et al.\ \cite{kaplan}).

Note that the cooling of neutron stars can
also be affected by the singlet-state superfluidity of neutrons in
inner stellar crusts,
by the presence of surface layers of light (accreted)
elements, and by stellar magnetic fields 
(e.g., Potekhin et al.\ \cite{pycg03},
Geppert et al.\ \cite{geppert}). 
These effects can be especially important in low-mass stars.
We have neglected them in the present paper  
since we have mainly focused on enhanced cooling
of massive stars but we will consider them
in a future publication.

%%%%%%%%%%%%%%%%%%%%%%%%%%%%%%%%%%%%%%%%%%%%%%%%%%%%%%%%
\section{Conclusions}
\label{concl}
%%%%%%%%%%%%%%%%%%%%%%%%%%%%%%%%%%%%%%%%%%%%%%%%%%%%%%%%

We have proposed a new scenario of cooling of
isolated neutron stars. We have shown that 
the present observational data on thermal emission
from isolated middle-aged neutron stars can be
explained assuming that neutron star cores are
composed of neutrons, protons and electrons
(and possibly muons) with forbidden direct Urca process
of neutrino emission. 
In our scenario, an enhanced neutrino emission,
which is required for interpretation of neutron stars
coldest for their age, is provided by neutrino process
associated with Cooper pairing of neutrons. We have shown
that the neutrino luminosity due to this process
(at internal temperatures $T \la 0.6\,T_{\rm cn}^{\rm max}$) 
behaves as $T^8$. In this way it ``mimics'' 
the neutrino luminosity produced either by modified Urca
processes or nucleon-nucleon bremsstrahlung processes in
non-superfluid stars, but it can be one-two orders
of magnitude higher. The proposed cooling scenario
imposes very stringent constraints on the density dependence
of neutron-pairing temperature
$T_{\rm cn}(\rho)$. The constraints 
result from the comparison
of the cooling theory with two most important
``testing sources'', PSR J0205+6449 and the Vela
pulsar (Sect.\ \ref{testing}). This scenario is the first
one in which a moderate superfluidity and associated
neutrino emission are helpful for explaining the data
(cf.\ with previous cooling scenarios, where a moderate
superfluidity has violated interpretation of observations,
e.g., Kaminker et al.\ \cite{kyg02}).

Our interpretation implies the presence of a strong
proton superfluidity and a moderate neutron superfluidity
in neutron star cores (Sect.\ \ref{physics}).
We need the proton superfluidity to explain observations
of neutron stars hottest for their age, and the neutron superfluidity
to explain observations of stars coldest for their age.
However, as has been demonstrated by
Gusakov et al.\ (\cite{gkyg04}), cooling curves 
are not too sensitive to 
exchanging neutron and proton superfluidities
($T_{\rm cp}(\rho)  
\rightleftharpoons
T_{\rm cn}(\rho)$)
in neutron-star cores. Therefore, we would also be able to
explain observational data in the scenario with a strong
neutron superfluidity and a moderate proton superfluidity
in stellar cores.  

We need a strong superfluidity to suppress modified Urca process
in low-mass stars, rise the surface temperature
of these stars and explain observations of neutron stars
hotter for their age. In fact, we can rise the temperature
of low-mass middle-aged neutron stars by assuming the
presence of surface layers of light (accreted) elements.
The mass of light elements may decrease with time, e.g., due
to diffusive nuclear burning (Chang \& Bildsten \cite{cb03}),
which opens additional freedom to regulate the cooling.
In this way, the presence of a strong (proton or neutron)
superfluidity in a neutron star core is not vitally
important for our interpretation. We will show this in a 
future publication. However, the presence of a moderate
superfluidity (of neutrons or protons) with a tuned
density dependence of critical temperature (Sect.\ \ref{testing})
is crucial for this scenario, where this tuned
dependence is combined with the remarkable simplicity
of the equation of state of neutron-star cores 
(nucleon composition with forbidden direct Urca process).
We hope that this scenario
can be taken into consideration along with many other
scenarios (reviewed or proposed, e.g., by
Page \cite{page98a,page98b},
Tsuruta et al.\ \cite{tsurutaetal02},
Khodel et al.\ \cite{khodeletal04},
Blaschke et al.\ \cite{blaschke},
Yakovlev \& Pethick \cite{yp04}).
The correct scenario should be selected in future
observations of neutron stars combined with
new advanced theoretical results.  

%%%%%%%%%%%%%%%%%%%%%%%%%%%%%%%%%%%%%%%%%%%%%%%%%%%%%%%%%%%%%
After this paper was prepared for submission we
became aware of the paper of Page et al.\ (\cite{pageetal04}).
These authors give a detailed consideration of enhanced cooling
via neutrino emission due to Cooper pairing
of neutrons in neutron-star cores composed of nucleons
with forbidden direct Urca process. 
The idea to enhance the cooling by Cooper-pairing neutrino emission
is the same as in our
paper, but its realization is different.
Particularly, Page et al.\ (\cite{pageetal04}) 
use a set of superfluidity models obtained from
microscopic theories. Their main models for neutron superfluidity
in a stellar core
(for, instance, model (a) in their Fig.\ 9) have too high
peak temperatures $T_{\rm cn}^{\rm max} \ga 10^9$ K and
too high $T_{\rm cn}(\rho)$ 
at the pre-peak densities to explain the observations of 
PSR J0205+6449 and the Vela pulsar and to obtain
a pronounced dependence of cooling curves on neutron
star mass. In contrast,
our $T_{\rm cn}(\rho)$ models are phenomenological
but tuning them we obtain a noticeable dependence of
the cooling on $M$. It enables us to
explain  all the data by one model of nucleon superfluidity
(even neglecting the effect of accreted envelopes).
 
\begin{acknowledgements}
We are grateful to A.Y.\ Potekhin for helpful discussions. 
This work has been supported partly by
the RFBR, grants 02-02-17668 and 03-07-90200,
the RLSS, grant 1115.2003.2,
and by the INTAS, grant YSF 03-55-2397.
\end{acknowledgements}

\end{document}